\documentclass[prc,twocolumn,nofootinbib]{revtex4-2}

\usepackage{color}
\usepackage{graphicx}
\usepackage{dcolumn}
\newcolumntype{d}[1]{D{.}{.}{#1}}
\usepackage{bm}
\usepackage{braket}
\usepackage{amsmath}
\usepackage{here}
\usepackage{color}
\usepackage{multirow}

\usepackage{array}
\usepackage{enumerate}
\usepackage{amssymb}
\usepackage{bbm}
\usepackage{dsfont}
\usepackage[scr=rsfs]{mathalpha}
\usepackage{longtable}

\def\be{\begin{equation}} 
\def\ee{\end{equation}}

\usepackage{xcolor}
\graphicspath{{figs/}}
\usepackage[bookmarks=true,colorlinks,
            citecolor=blue,linkcolor=blue,anchorcolor=blue,filecolor=blue,urlcolor=blue,
           ]{hyperref}

\usepackage[normalem]{ulem}

\renewcommand\sout{\bgroup\markoverwith
{\textcolor[rgb]{1,0.75,0.8}{\rule[.5ex]{2pt}{0.8pt}}}\ULon}

\begin{document}

\title{
Re-examination of fusion hindrance in astrophysical 
$^{12}$C+$^{12}$C and $^{12}$C+$^{13}$C reactions
}

\author{K. Uzawa}
\affiliation{ 
Nuclear Data Center, Japan Atomic Energy Agency, Tokai 319-1195,  Japan} 

\author{K. Hagino}
\affiliation{ 
Department of Physics, Kyoto University, Kyoto 606-8502,  Japan} 
\affiliation{Institute for Liberal Arts and Sciences, Kyoto University, Kyoto 606-8501, Japan}
\affiliation{ 
RIKEN Nishina Center for Accelerator-based Science, RIKEN, Wako 351-0198, Japan
}

\begin{abstract}
To determine the energy dependence of fusion cross sections at extremely low energies is crucial 
for various astrophysical processes. 
In the previous study by Jiang et al. [Phys. Rev. C75, 015803 (2007)], it was concluded that fusion cross sections 
for the $^{12}$C+$^{12}$C system rapidly drop off as the energy decreases. We here re-examine this hindrance phenomenon. 
While the previous study fitted the logarithmic slope $L(E)$ of fusion cross sections with a function 
of $L(E)=A+B/E^n$ and searched the optimum value of $A$ and $B$ with $n=1.5$, we refit the data with the same 
function for $L(E)$ but by releasing 
the restriction on $n$. We find that  the optimum values of $n$ significantly deviates from $n=1.5$, resulting 
in the absence of hindrance of fusion cross sections both in the $^{12}$C+$^{12}$C 
and the $^{12}$C+$^{13}$C 
systems. 
\end{abstract}

\maketitle

{\it Introduction.}
The $^{12}$C+$^{12}$C fusion reaction is one of the most important nuclear astrophysical reactions. 
This reaction plays an essential role in carbon burning in massive stars, type Ia supernovae, and X-ray 
superburst. It is therefore crucially important to have a reliable extrapolation of the fusion cross 
sections, or the nuclear astrophysical $S$-factors, down to the Gamow energy.  

An important issue which has been under debate is whether there is the fusion hindrance 
in the $^{12}$C+$^{12}$C reaction at astrophysical energies. The fusion hindrance was first recognized 
in medium-heavy systems, such as $^{60}$Ni+$^{89}$Y system \cite{Jiang2002},  
for which the measured fusion cross sections fall off more steeply at energies well below the barrier 
than theoretical calculations 
based on the coupled-channels approach \cite{back2014,Jiang_review,montagnoli2017}.
In Ref. \cite{Jiang2007a}, Jiang et al. analyzed the logarithmic derivative of fusion cross sections 
for light systems including 
$^{12}$C+$^{12}$C and $^{16}$O+$^{16}$O and argued that there is a strong hindrance phenomenon in 
such light systems as well. 
This is crucially important for nuclear astrophysics, since the superburst would not be ignited 
if the hindrance phenomenon exists \cite{cooper2009}. 

The experimental situation has not been clear concerning the presence or the absence of the hindrance 
phenomenon in the $^{12}$C+$^{12}$C fusion reaction. 
Recent direct measurements of fusion cross sections are not inconsistent with the prediction of the 
hindrance model \cite{fruet2020,tan2020}, while another direct measurement in Ref. \cite{Spillane2007} 
does not disentangle the situation because of the experimental uncertainties. 
On the other hand, the hindrance model was supported neither by the indirect 
method with the Trojan Horse method (THM) \cite{Tumino2018} nor by the measurements for the 
$^{12}$C+$^{13}$C system \cite{Zhang2020}, even though the astrophysical $S$ factor from the THM may be 
consistent with the hindrance model when the Coulomb correction is introduced \cite{Mukhamedzhanov2019,Beck2020}. 
Moreover, recent theoretical calculations based on the anti-symmetric molecular dynamics are also consistent with the 
absence of the fusion hindrance in $^{12}$C+$^{12}$C \cite{taniguchi2021,taniguchi2024}. 

Give this situation, it is our aim in this paper to 
reexamine the hindrance phenomenon in the $^{12}$C+$^{12}$C system. 
In Ref. \cite{Jiang2007a}, the experimental logarithmic slope $L(E)$ of fusion cross sections $\sigma$, 
\begin{equation}
    L(E)=\frac{d}{dE}{\rm ln}(E\sigma(E))=\frac{1}{E}+\frac{1}{\sigma(E)}\,\frac{d\sigma(E)}{dE},
    \label{L_def}
\end{equation}
where $E$ is the incident energy in the center of mass frame, was fitted 
with a function 
\begin{equation}
    L(E)=A+B/E^n,
\label{L(E)}
\end{equation}
where $A,B,$ and $n$ are adjustable parameters. 
Motivated by the expression of $L(E)$ for a constant astrophysical $S$ factor, the authors of Ref. \cite{Jiang2007a} 
fixed the value of $n$ to be 1.5, and obtained a parameter set of $A$ and $B$ which leads to a strong hindrance of fusion cross sections 
at extremely low energies. 
However, for several reasons astrophysical $S$ factors can deviate from a constant value for several reasons, and 
it is not necessary to expect that the astrophysical $S$-factors follow the same energy dependence as a constant $S$-factor. 
In this paper, we thus release the restriction of $n=1.5$ and refit the experimental data by adjusting 
all of $A,B$, and $n$. Notice that $n$ is directly connected to the energy dependence of $L(E)$ and thus 
the hindrance phenomenon 
is sensitive largely to the value of $n$. Together with the $^{12}$C+$^{12}$C system, we shall also analyze the 
$^{12}$C+$^{13}$C system. 
The $^{12}$C+$^{12}$C reaction shows large fluctuations of the reaction rate 
around $E\leq 4$ MeV due to resonances, which may deteriorate the simple form of $L(E)$ given by Eq. (\ref{L(E)}). 
On the other hand, the cross sections of the $^{12}$C+$^{13}$C system 
show a much smoother energy dependence, and thus the extrapolation of cross sections to low energies would be more reliable. 
In addition, we shall estimate the uncertainties of the parameter fitting and their 
propagation to $S(E)$ and $L(E)$. 
Such error estimation is crucial particularly 
when one discusses the hindrance phenomenon in the energy region in which experimental data are scarce.

{\it Fitting procedure. }
Using the fusion cross section $\sigma(E)$,
the astrophysical $S$-factor is defined as
\begin{equation}
    S(E)=\sigma(E)E\exp(2\pi\eta(E)). 
\end{equation}
Here, $\eta$ is the Sommerfeld parameter defined as 
$\eta(E)=Z_1Z_2e^2/\hbar v$, where $Z_1$ and $Z_2$ are the atomic number of the projectile and the target nuclei, respectively,  
and $v=\sqrt{2E/\mu}$ with $\mu$ being the reduced mass is the classical velocity for the relative motion between the colliding nuclei. 
For collisions between two point charges,  
$S(E)$ is a constant and the logarithmic slope $L(E)$ reads \cite{Jiang2004, Jiang2007a, Jiang2007b,hagino2003},  
\begin{equation}
    L_{cs}(E)=\frac{\pi\eta}{E}. 
    \label{Lcs}
\end{equation}
Notice that the astrophysical $S$ factor $S(E)$ and the logarithmic slope $L(E)$ are related to each other 
as 
\begin{equation}
    \frac{dS}{dE}=S(E)\left[L(E)-\frac{\pi\eta}{E}\right]=S(E)\left[L(E)-L_{cs}(E)\right], 
\end{equation}
implying that 
$S(E)$ takes a maximum at $E=E_s$, at which $L(E)$ intersects with $L_{cs}(E)$, that is, $L(E_s)=L_{cs}(E_s)$. 
Below $E_s$, $S(E)$ starts decreasing as $E$ decreases, that has been regarded as a signature of 
the fusion hindrance phenomenon \cite{Jiang_review,hagino2018}. 

Notice that for a system with a positive reaction $Q$ value, the fusion cross section is finite at $E=0$, 
and both $L(E)$ and $L_{cs}$ diverge at this energy (see Eqs. (\ref{L_def}) and (\ref{Lcs})).
Therefore, for such systems, it is not guaranteed that $L$ intersects with $L_{cs}$ and thus the hindrance 
behavior appears. 
In the following, we assumes that $L(E)$ can be well approximated by the function (\ref{L(E)}). 
When $n$ takes a positive value, $L(E)$ diverges as $E\to 0$, 
and the condition for a positive $Q$ value is satisfied. 
On the other hand, 
for a negative value of $n$, $L(E)$ approaches $A$ as $E\to 0$, and the condition is not met. 
Therefore a positive $n$ is preferable, even though 
a negative value of $n$ cannot be excluded if the functional form is not reasonable outside the fitting region. 
If $n$ is negative, $n$ should be regarded as an effective value that is valid in the limited energy region. 


%


In the previous work \cite{Jiang2007a},
the parameters $A$ and $B$ in Eq. (\ref{L(E)}) were determined by fitting to the 
experimental data for $L(E)$.
However, 
$L(E)$ is not a direct observable but is obtained by taking the energy derivative. 
Due to wide and non-uniform energy spacings as well as the experimental uncertainties in the data, 
the resultant values of $L(E)$ can be affected by a numerical scheme employed 
to take the energy derivative. 
We therefore fit the experimental data for the astrophysical $S$ factor, rather than the logarithmic 
slope $L(E)$, using the function,
\begin{equation}
S(E) = \sigma' E' e^{C(E)},
\label{S_int}
\end{equation}
with
\begin{equation}
C(E)=A(E-E')- B\frac{\left( \frac{E'}{E} \right)^{n-1} - 1}{E'^{n-1}(n-1)}  + 2\pi\eta(E), 
\label{S_int2}
\end{equation}
which is obtained by integrating Eq. (\ref{L(E)}) \cite{Jiang2007a}.
Here, $E'$ is the lower limit of the integration,
\begin{equation}
    \int^E_{E'} dE''\frac{d}{dE''}{\rm ln}(E''\sigma(E''))={\rm ln}(E\sigma(E))-{\rm ln}(E'\sigma(E')),
\end{equation}
and $\sigma'$ is the cross section at $E'$, that is, $\sigma'=\sigma(E')$. 
Notice that $E'$ can be chosen arbitrarily and in this paper we take $E'=2$ MeV. 
$\sigma'$ is determined
together with $A, B$ and $n$ in the least squares fitting.

In the least squares fitting, 
we employ both the weighted least squares (WLS) method and the ordinary least squares (OLS) method. 
In the WLS, 
the chi-square is defined as 
\begin{equation}
    \chi^2_w\equiv\sum_i \left( \frac{S_i-S(E_i)}{\Delta S_i} \right)^2,
\end{equation}
where $S_i$ and $\Delta S_i$ are the $S$ factor and its experimental uncertainty, respectively,  
at the $i$-th experimental data point, $E_i$. 
$S$ is the fitting function given by Eqs. (\ref{S_int}) and (\ref{S_int2}). 
On the other hand, the chi-square in OLS is defined as,  
\begin{equation}
    \chi^2_o=\sum_i \left( \frac{S_i-S(E_i)}{\Delta S_0}\right)^2,
\end{equation}
where $\Delta S_0$ is the estimated standard deviation defined by 
\begin{equation}
   \Delta S_0=\sqrt{\frac{1}{N-m}\sum_i (S_i-S_{\rm best}(E_i))^2}. 
\end{equation}
Here $S_{\rm best}$ is the best-fit function of $S(E)$, $N$ is the number of the data-points, and $m$ is the number of fitting parameters. The actual value of $\Delta S_0$ does not affect the estimated values of the parameters, even 
though its value is necessary to determine the covariance matrix of the fitting parameters.

Usually, the OLS method is employed when the experimental errors are not known. 
However, 
the experimental $S$ factors have a larger uncertainty in the low-energy region as compared to those in the 
high-energy region, and a little weight is given to the data in the low-energy region when the WLS method is applied. 
Therefore, the OLS method may work better in the present case where the energy dependence of $S$ factor in the 
low-energy region is discussed, as it treats all the data points equally. 

After the parameters are optimized with the chi-square fittings, one can compute 
the covariance of the parameters defined as \cite{bevington2003data}, 
\begin{equation}
\label{cov}
    {\rm Cov}^{-1}(x_i,x_j)=\left.\frac{1}{2}\frac{\partial^2\chi^2}{\partial x_i\partial x_j}\right|_{\vec{x}'},
\end{equation}
where $\vec{x}=(x_1,x_2,...,x_m)$ are the fitting parameters and 
$\vec{x}'$ is the set of the best-fit parameters.
Using the covariance, the variance of the function $S(E)$ reads \cite{bevington2003data} 
\begin{equation}
    {\rm Var}(S(E))=\left.\sum_{i,j}\frac{\partial S(E)}{\partial x_i}\right|_{_{\vec{x}'}}{\rm Cov}(x_i,x_j)
    \left.\frac{\partial S(E)}{\partial x_j}\right|_{_{\vec{x}'}}.
    \label{VofS}
\end{equation}
Based on Eq. (\ref{VofS}), we shall discuss the uncertainty of the predicted $S$-factor. 
The variance of $L(E)$ is calculated in the same way, except that 
the contribution of the parameter $\sigma'$ is suppressed.

\begin{figure}[tb]
  \centering
  \includegraphics[width=8.6cm]{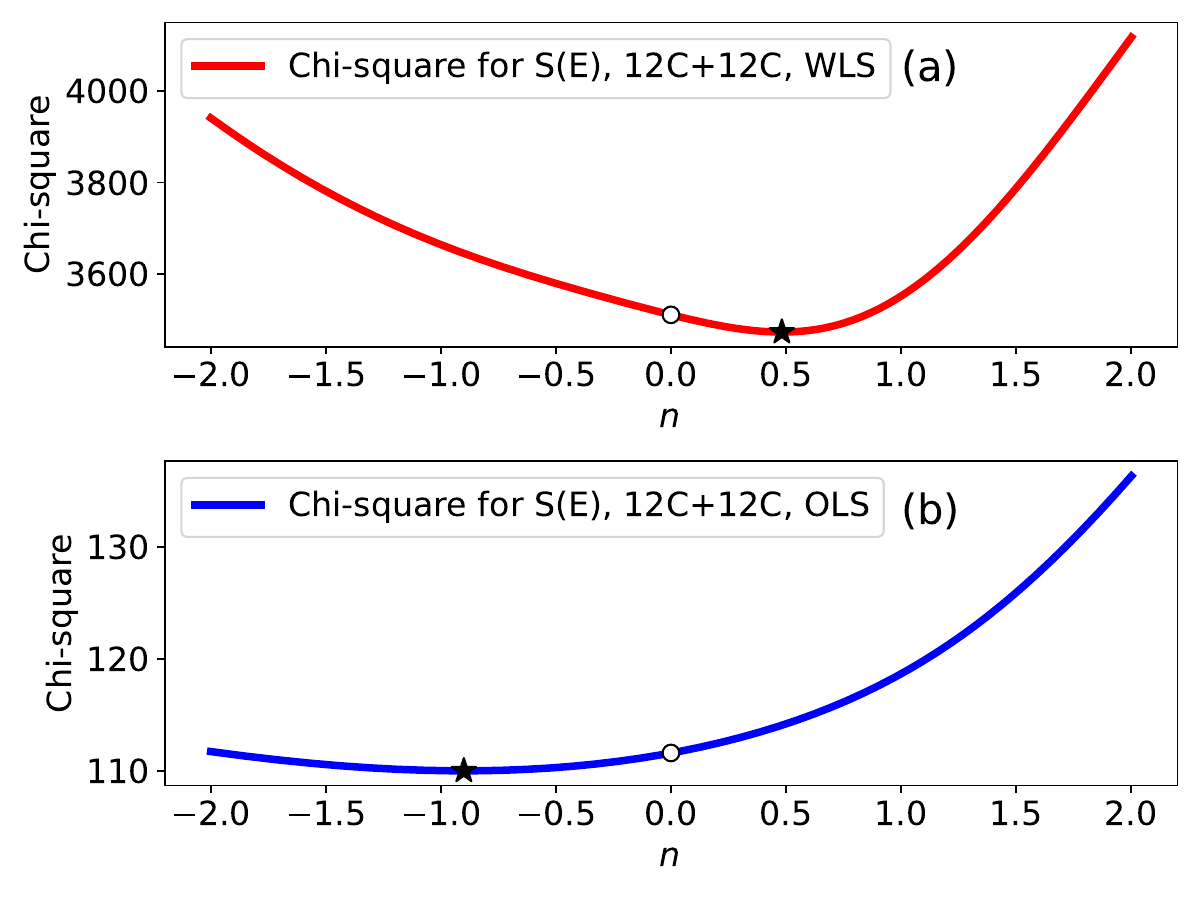} 
  \caption{
  (a) The chi-square values $\chi^2_w$ for the WLS fitting to the $S$ factor for the $^{12}$C+$^{12}$C fusion reaction. 
  Here, the value of $n$ is fixed, and the other parameters $A, B$ and $\sigma'$ in Eq. (\ref{S_int}) are optimized for each $n$. 
  The fitting excludes $n=0$. The star shows the minimum of the chi-square.
  (b) Same as (a), but for the OLS fitting. 
  }
  \label{fig1}
\end{figure}

{\it Results. }
Let us now apply the fitting procedures to 
the $^{12}$C+$^{12}$C and $^{12}$C+$^{13}$C reactions. 
We first discuss the $^{12}$C+$^{12}$C system. 
For this reaction, we adopt the experimental data 
from Refs. \cite{Patterson1969, Mazarakis1973, High1977, Aguilera2006, Jiang2018b}, three of which 
\cite{Patterson1969, Mazarakis1973, High1977} were used in Ref. \cite{Jiang2007a} while we add the remaining two \cite{Aguilera2006, Jiang2018b} to reduce the uncertainty of the fitting \footnote{
Besides these references, Refs.\cite{Barron2006,Spillane2007, Zickefoose2011} also 
report the measurements of fusion cross sections for this system in the low-energy region. 
Even though these provide important information on the low-energy behavior of the $^{12}$C+$^{12}$C reactions, 
we do not adopt them in this paper as they systematically deviate from the other adopted data.}. 
Figure \ref{fig1} shows the chi-square values for the WLS and the OLS fittings to the astrophysical $S$ factors.  
Here the value of $n$ in Eq. (\ref{S_int}) is fixed and the remaining three parameters, 
$A$, $B$ and $\sigma'$ are optimized for each $n$. 
We exclude the point of $n=0$ from the figure, 
as $L(E)$ becomes a constant at this point and physically reasonable results cannot be obtained. 

\begin{table}[tb]
\centering
\caption{
The best-fit parameters and their uncertainties for the weighted least squares (WLS) and the ordinary least squares (OLS) 
fittings 
to the experimental astrophysical $S$ factors for the $^{12}$C+$^{12}$C system with the function given by Eq. (\ref{S_int}). 
}
\label{opt_12C}
\begin{tabular}{c|>{\centering\arraybackslash}p{3.1cm}|>{\centering\arraybackslash}p{2.9cm}}
\hline
\hline
        & WLS           & OLS\\
\hline
$A \ {\rm (MeV^{-1})}$ & $-12.52\pm1.83$    & $13.47\pm6.30$   \\
$B \ (\mathrm{MeV}^{\mathit{n}-1})$ & $34.61\pm0.96$    & $-2.37\pm4.22$   \\
$\sigma' \ {\rm (barn)}$ & $\begin{array}{c}
(1.67\pm0.25)\times10^{-11} \\
\end{array}$  & $\begin{array}{c}
(1.74\pm0.46)\times10^{-10} \\
\end{array}$  \\
$n$ & $0.482\pm0.055$    & $-0.90\pm0.73$   \\
\hline
\hline

\end{tabular}
\end{table}

The best-fit parameters are summarized in Table \ref{opt_12C}.
The errors of those parameters are  estimated from the diagonal components of the covariance matrix given by Eq. (\ref{cov}).
The best-fit value of $n$ is $n=0.48$ for the WLS fit and $n=-0.90$ for the OLS fit, both of which 
deviate significantly from $n=1.5$ adopted in Ref. \cite{Jiang2007a}. 
The variances of $n$ is $2.97\times10^{-3}$ and $0.53$ in the WLS and OLS fits, respectively, 
and $n=1.5$ are excluded within the range of $1\sigma$ in both cases.
The relatively large variance in the OLS case reflects a shallow minimum of the chi-square shown 
in Fig. \ref{fig1}(b).

\begin{figure}[htbp]
  \centering
  \includegraphics[width=0.8\columnwidth]{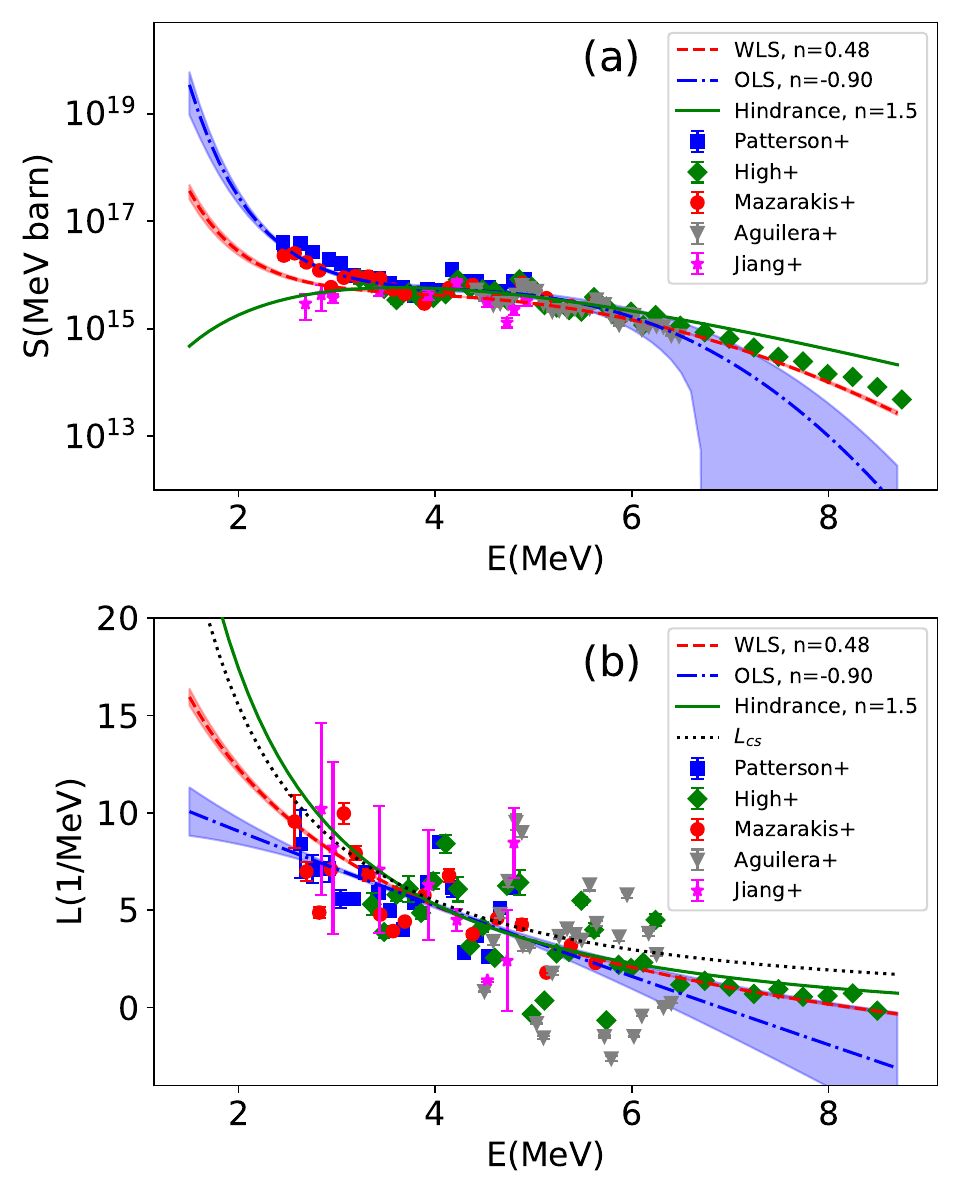} 
  \caption{(a) The astrophysical $S$ factors for the $^{12}$C+$^{12}$C system calculated with 
  Eqs. (\ref{S_int}) and  (\ref{S_int2}) 
  obtained with the parameters the WLS (the red dashed line) and the OLS (the blue dot-dashed line) 
  fittings, respectively. The $S$ factors from the hindrance model of Ref. \cite{Jiang2007a} are also shown for comparison by the 
  green solid line. The experimental data are taken from  Refs. \cite{Patterson1969, Mazarakis1973, High1977, Aguilera2006, Jiang2018b}. 
(b) The corresponding logarithmic slopes, $L(E)$. The logarithmic slope with a constant $S$ factor, $L_{cs}$, is also 
shown by the dotted line.}
  \label{fig2}
\end{figure}

The astrophysical $S$ factor calculated with these best-fit parameters are plotted in the upper panel of Fig. \ref{fig2}.
The error-range is calculated based on 
Eq. (\ref{VofS}) with the obtained covariance matrices, 
which are summarized in the Supplemental Material.  
For comparison, the figure also shows the astrophysical $S$ factor from the hindrance model of Ref. \cite{Jiang2007a}. 
Because the WLS fit puts an emphasis on the data points with small errors,
the fit shows a good agreement with the experimental data in the high energy region (see the red dashed line).
In contrast,
the OLS fit leads to a better agreement in the lower energy region around $2 \leq E\leq 4$ MeV by 
reducing the value of $n$ (see the blue dot-dashed line). 
In any case, it is remarkable that the astrophysical $S$ factor obtained with the best-fit parameters 
does not drop off as the energy decreases, and thus the hindrance phenomenon does not occur in the $^{12}$C+$^{12}$C reaction as long as Eqs. (\ref{S_int}) and (\ref{S_int2}) are employed. 

The corresponding logarithmic slope $L(E)$ is shown in Fig. \ref{fig2} (b). The error band is estimated in a similar way 
to that for the astrophysical $S$ factor. 
The energy derivative in $L(E)$ is calculated by fitting the three neighboring points of ${\rm ln}(E\sigma)$ by a straight line. 
One can see that the small value of $n$ in the OLS fit suppresses $L(E)$ in the low energy region.
One can also notice that the resultant logarithmic slopes do not intersect with 
$L_{cs}$ shown by the dotted line. This is consistent with the absence of the hindrance phenomenon 
demonstrated in the upper panel of the figure. 

\begin{figure}[htbp]
  \centering
  \includegraphics[width=8.6cm]{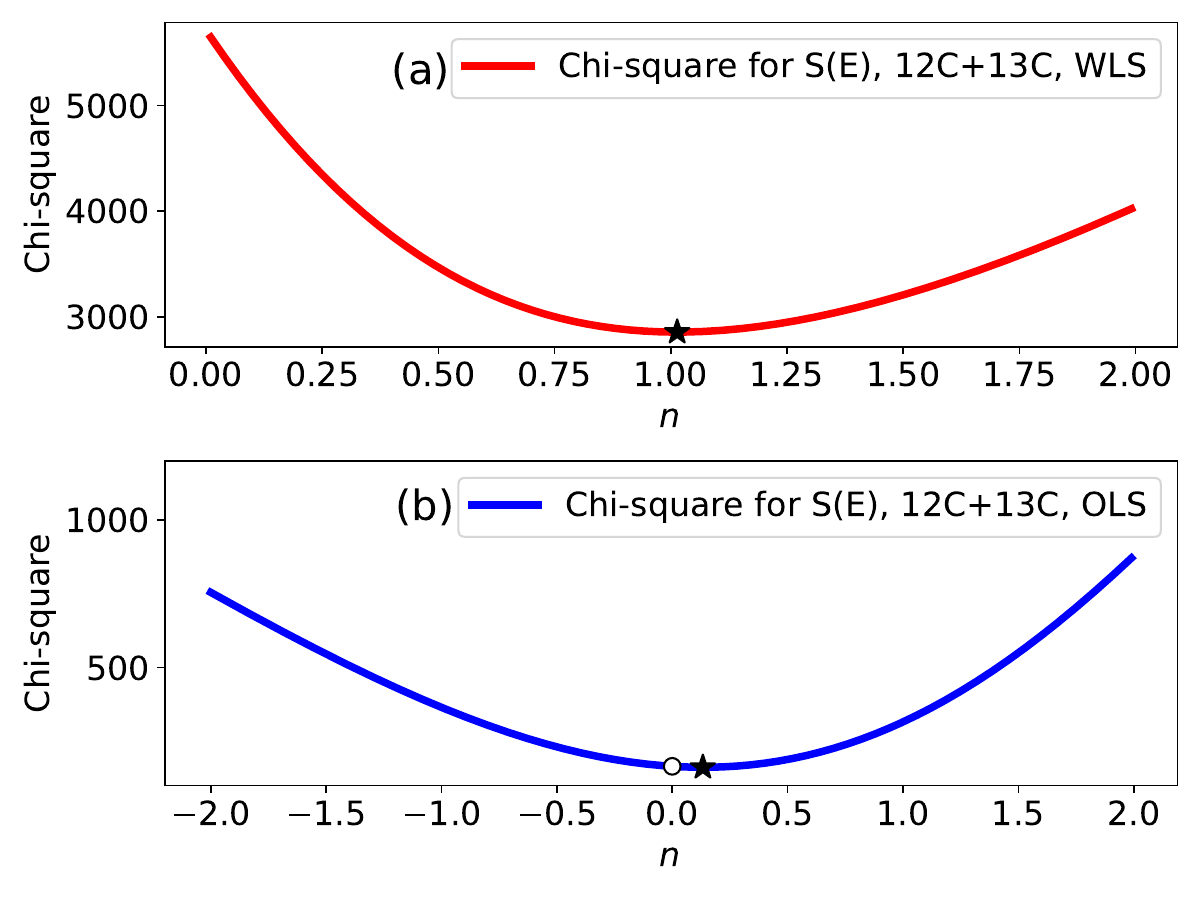} 
  \caption{Same as Fig. \ref{fig1}, but for the $^{12}$C+$^{13}$C fusion reaction.
  }
  \label{fig3}
\end{figure}

Let us next discuss the $^{12}$C+$^{13}$C system.
To this end, we adopt the experimental data in Refs. \cite{Stokstad1976, Dasmahapatra1982, ND2012, Zhang2020}, 
except for the highest energy point in Ref. \cite{Stokstad1976}, at which the validity of 
Eq. (\ref{L(E)}) may be questionable. 
See the Supplemental Material for the dependence of the result on the energy cutoff in the experimental data. 
Figures \ref{fig3}(a) and (b) show the chi-square functions, $\chi^2_w$ and $\chi^2_o$, for the 
WLS and the OLS fittings to the astrophysical $S$ factors with different values of $n$.
The resultant values of $n$ are $n=1.01$ and $n=0.13$ for the WLS and the OLS fittings, respectively.
Contrary to the $^{12}$C+$^{12}$C case, 
the resultant values of $n$ are positive in both cases.
One can notice that the values of $n$ deviate from $n=1.5$, as for the $^{12}$C+$^{12}$C reaction. 
See Table \ref{opt_13C} for the best-fit values for the other parameters. 

\begin{table}[tb]
\centering
\caption{Same as Table \ref{opt_12C}, but for  the $^{12}$C+$^{13}$C reaction. 
}
\label{opt_13C}
\begin{tabular}{c|>{\centering\arraybackslash}p{3.1cm}|>{\centering\arraybackslash}p{2.9cm}}
\hline
\hline
        & WLS           & OLS\\
\hline
$A \ {\rm (MeV^{-1})}$  & $-5.49\pm0.19$    & $-74.48\pm43.45$   \\
$B \ (\mathrm{MeV}^{\mathit{n}-1})$ & $43.94\pm0.62$    & $96.10\pm42.69$   \\
$\sigma' \ {\rm (barn)}$ &  $\begin{array}{c}
(1.21\pm0.01)\times10^{-11} \\
\end{array}$   &  $\begin{array}{c}
(4.66\pm0.33)\times10^{-12} \\
\end{array}$    \\
$n$ & $1.01\pm0.02$    & $0.132\pm0.072$   \\
\hline
\hline

\end{tabular}
\end{table}

\begin{figure}[htbp]
  \centering
  \includegraphics[width=8.6cm]{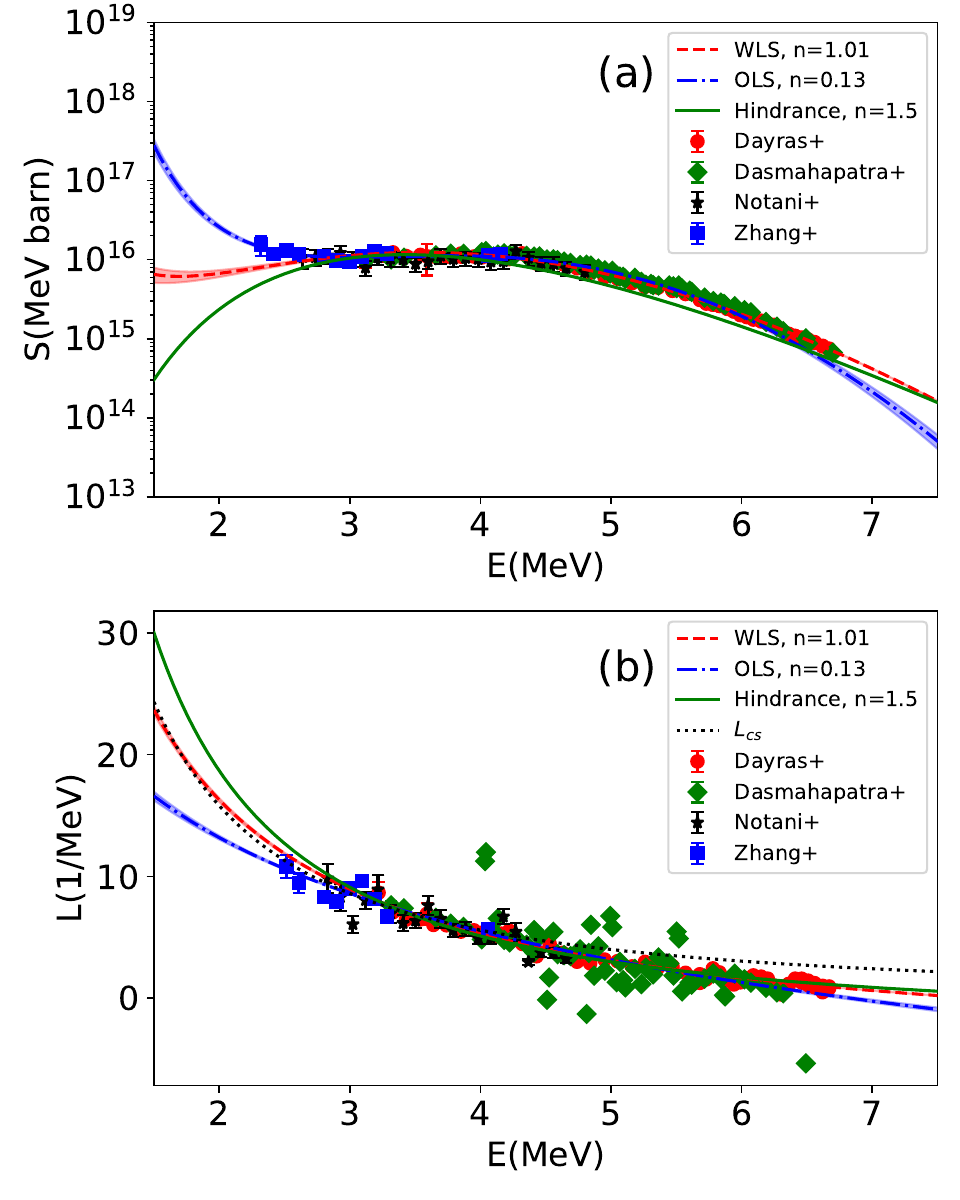} 
  \caption{Same as Fig. \ref{fig2}, but for the $^{12}$C+$^{13}$C fusion reaction.}
  \label{fig4}
\end{figure}

\begin{figure}[htbp]
  \centering
  \includegraphics[width=8.6cm]{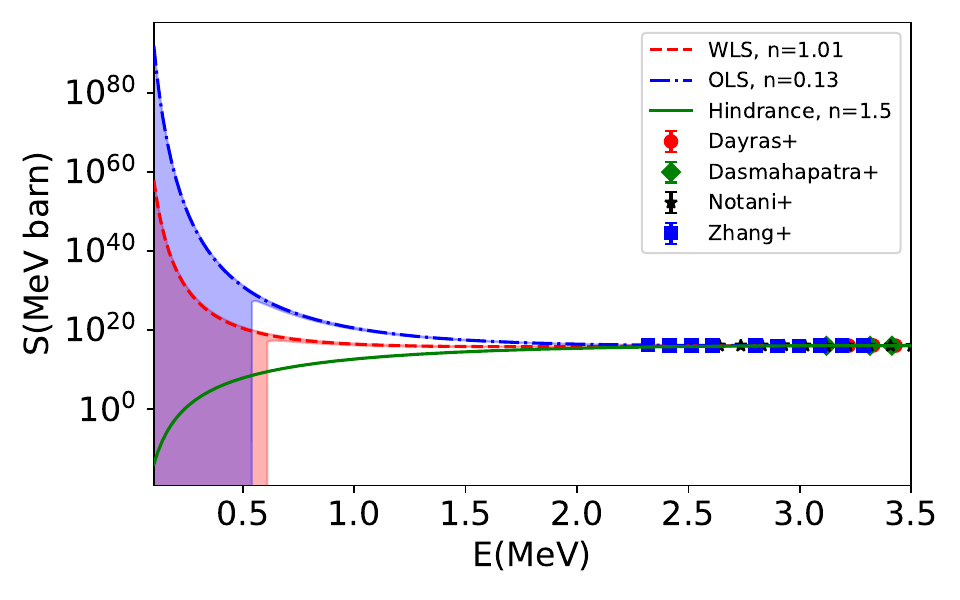} 
  \caption{Same as Fig. \ref{fig4}(a) but with the expanded scale for the vertical axis.}
  \label{fig5}
\end{figure}

The astrophysical $S$ factors are shown in Fig. \ref{fig4}(a).
Similarly to Fig. \ref{fig2} (a), 
the WLS fit emphasizes the agreement to the data in the high energy region,
and the OLS fit leads to a better fit in the lower energy region below 3 MeV.
One can clear see that the hindrance does not take place in this reaction (see Fig. \ref{fig5} for the 
expanded scale of Fig. \ref{fig4}(a)). 
This is consistent with the corresponding logarithmic slopes $L(E)$ shown in Fig. \ref{fig4}(b).
Even though the WLS results intersects with $L_{cs}$, the $S$ factor at the maximum energy is rather flat and 
thus no clear hindrance appears. In the case of the OLS fitting, $L(E)$ does not intersect with $L_{cs}$ due 
to the small $n$. 
Notice that, in both the WLS and the OLS cases, the error ranges shown in Fig.\ref{fig4} are smaller, 
and thus the fitting is more reliable,  
compared to those in Fig. \ref{fig2} for the $^{12}$C+$^{12}$C reaction. 
In the  $^{12}$C+$^{12}$C reaction, the prominent resonance structures appear in the low energy region,
which lead to larger uncertainties in the fitting as compared to the $^{12}$C+$^{13}$C reaction. 
This is also reflected in the values of the covariance matrices shown in the Supplemental Materials.

{\it Summary. }
We have reexamined 
the fusion hindrance phenomenon in the $^{12}$C+$^{12}$C and $^{12}$C+$^{13}$C reactions. 
While the previous studies fitted the logarithmic slopes $L(E)$ with the function 
$L(E)=A+B/E^n$ by fixing $n=1.5$, we varied the value of $n$ as well. 
We have found that both the 
weighted least squares (WLS) and the ordinary least squares (OLS) 
fittings lead to the optimized value of $n$ which is significantly different from $n=1.5$. 
We have shown that the resultant astrophysical $S$ factors do not show the hindrance behaviour 
within the range of error bars for 
both the $^{12}$C+$^{12}$C and the $^{12}$C+$^{13}$C systems.
Even though
the result of a fitting may depend on a selection of data points as well as a fitting scheme, 
it is evident that 
the two standard fitting methods, WLS and OLS, do not support the hindrance scenario in 
the C+C fusion reactions, at least within the fitting function which we employed.  
This implies the necessity of a careful treatment of the hindrance phenomenon, including a better 
fitting function.

{\it Acknowledgments.}
We thank X.D. Tang and M. Kimura, and O. Iwamoto 
for useful discussions. 
This work was supported in part by
JSPS KAKENHI Grant Number JP23K03414.

\bibliography{references} 
\end{document}


\setcounter{table}{2}
\setcounter{figure}{5}

\section{Data of Covariances}
The covariance matrices obtained in the WLS and OLS fittings are summarized in Tables \ref{cov_12C_WLS}-\ref{cov_13C_OLS}.
These values are used for the calculations of the error ranges in Figs. 3, 4 and 5 in the main text.

\begin{table}[htbp]

\centering
\caption{The covariance matrix for the WLS fitting of the parameters $A$, $B$, $\sigma'$ and $n$ for the $^{12}$C+$^{12}$C reaction.
The units of the parameters $A$, $B$ and $\sigma'$ are ${\rm MeV^{-1}}$, $\mathrm{MeV}^{\mathit{n}-1}$ and ${\rm barn}$, respectively.
}
\label{cov_12C_WLS}
\begin{tabular}{c|cccc}
\hline
\hline
        & $A $            & $B $           & $\sigma' $                & $n$ \\
\hline
$A $  & 3.34   & $-$1.73   & $-3.97\times10^{-12}$  & $9.94\times10^{-2}$ \\
$B $ & $-$1.73   & 0.92   & $1.88\times10^{-12}$   & $-5.11\times10^{-2}$ \\
$\sigma' $  & $-3.97\times10^{-12}$ & $1.88\times10^{-12}$ & $5.99\times10^{-24}$ & $-1.22\times10^{-13}$ \\
$n$ &$9.94\times10^{-2}$ & $-5.11\times10^{-2}$ & $-1.22\times10^{-13}$ & $2.97\times10^{-3}$  \\
\hline
\hline
\end{tabular}
\end{table}

\begin{table}[htbp]

\centering
\caption{Same as Table \ref{cov_12C_WLS}, but for the OLS fitting.
}
\label{cov_12C_OLS}
\begin{tabular}{c|cccc}
\hline
\hline
        & $A$           & $B$           & $\sigma'$                & $n$ \\
\hline
$A$ & 39.67    & $-$26.56   & $-2.60\times10^{-10}$  & 4.55 \\
$B$ & $-$26.56   & 17.82    & $1.71\times10^{-10}$   & $-$3.06 \\
$\sigma'$ & $-2.60\times10^{-10}$ & $1.71\times10^{-10}$ & $2.10\times10^{-21}$ & $-2.89\times10^{-11}$ \\
$n$ & 4.55      & $-$3.06     & $-2.89\times10^{-11}$ & 0.53 \\
\hline
\hline

\end{tabular}
\end{table}

\begin{table}[htbp]

\centering
\caption{Same as Table \ref{cov_12C_WLS}, but for the $^{12}$C+$^{13}$C fusion reaction.
}
\label{cov_13C_WLS}
\label{tab:covariance4}
\begin{tabular}{c|cccc}
\hline
\hline

        & $A$           & $B$           & $\sigma'$                & $n$ \\
\hline
$A$ & $3.61\times10^{-2}$  & $1.15\times10^{-1}$  & $-2.14\times10^{-14}$  & $4.26\times10^{-3}$ \\
$B$ & $1.15\times10^{-1}$  & $3.89\times10^{-1}$  & $-7.52\times10^{-14}$  & $1.39\times10^{-2}$ \\
$\sigma'$ & $-2.14\times10^{-14}$ & $-7.52\times10^{-14}$ & $1.49\times10^{-26}$ & $-2.62\times10^{-15}$ \\
$n$ & $4.26\times10^{-3}$  & $1.39\times10^{-2}$  & $-2.62\times10^{-15}$  & $5.08\times10^{-4}$ \\
\hline
\hline

\end{tabular}
\end{table}

\begin{table}[htbp]

\centering
\caption{Same as Table \ref{cov_12C_OLS}, but for the $^{12}$C+$^{13}$C fusion reaction. 
}
\label{cov_13C_OLS}
\begin{tabular}{c|cccc}
\hline
\hline

         & $A$           & $B$           & $\sigma'$                & $n$ \\
\hline
$A$ & $1.89\times10^{3}$ & $-1.86\times10^{3}$ & $-1.20\times10^{-11}$  & $3.12$ \\
$B$ & $-1.86\times10^{3}$ & $1.82\times10^{3}$ & $1.17\times10^{-11}$  & $-3.06$ \\
$\sigma'$ & $-1.20\times10^{-11}$ & $1.17\times10^{-11}$ & $1.06\times10^{-25}$ & $-1.98\times10^{-14}$ \\
$n$ & $3.12$ & $-3.06$ & $-1.98\times10^{-14}$ & $5.14\times10^{-3}$ \\
\hline
\hline

\end{tabular}
\end{table}

\section{Selection of the data point}
\label{Selection of the data point}
\begin{figure}[htbp]
  \centering
  \includegraphics[width=8.6cm]{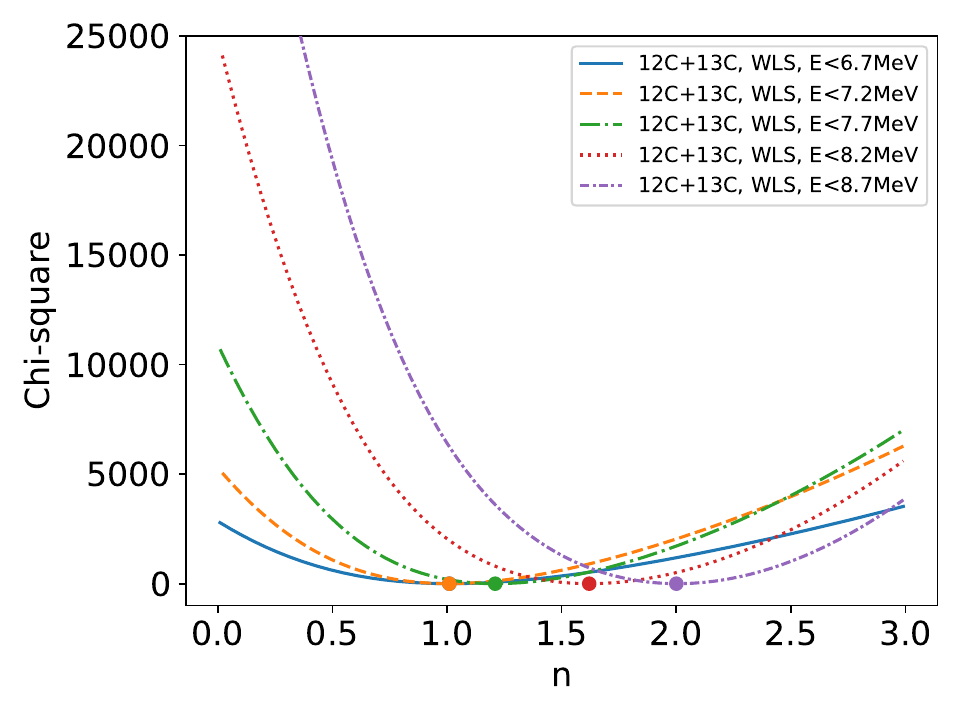} 
  \caption{
The relative values of chi-squares $\chi^2_w$ for 
the WLS fitting to the astrophysical $S$ factors for 
the $^{12}$C+$^{13}$C system 
with different values of the energy cutoff for the data points. }
  \label{fig6}
\end{figure}

\begin{figure}[htbp]
  \centering
  \includegraphics[width=8.6cm]{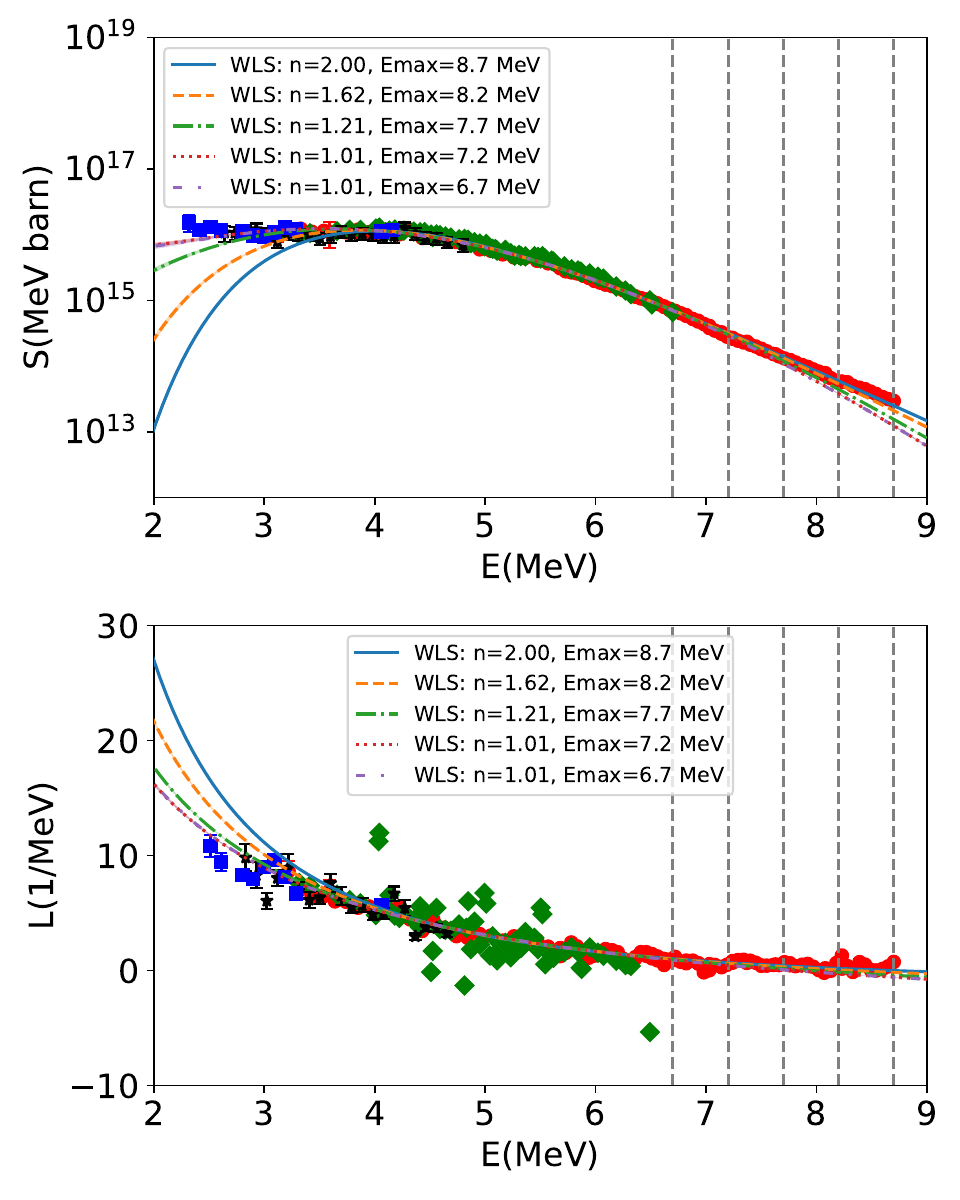} 
  \caption{Same as Fig. 4 in the main text, but with different maximum energies for the selected data points.
  The vertical lines indicate the energy cutoffs, 6.7, 7.2, 7.7, 8.2, and 8.7 MeV employed for each line.}
  \label{fig7}
\end{figure}

The fitting results are in general affected by how data points are selected.
In the analysis of $^{12}$C+$^{13}$C system in the main text, we set the upper limit of the energy to be 6.7 MeV.
Even though 6.7 MeV is not far from the fusion barrier height for this sytem, that is, around 5.7 MeV, 
some ambiguity may arise from the selection of the energy cutoff.
We find that the optimal value of $n$ is somewhat altered by the upper energy limit of the selected data, 
especially for the WLS fitting. To demonstrate this, 
Fig. \ref{fig6} shows the chi-square values $\chi^2_w$ 
for the WLS fitting to the astrophysical $S$ factors for the $^{12}$C+$^{13}$C reaction 
with different values of the upper limit of $E$.
One can see that the optimum value of $n$ tends to increase as the energy cutoff increases.
The mechanism for this is explained in Fig. \ref{fig7}.
The larger $n$ results in a good agreement in the higher energy region, larger than 7 MeV.
On the other hand, the smaller $n$ shows a good agreement in the lower energy region.
That is, the optimal value of $n$ is changed by a valance between the low-energy region 
and the high-energy region.
This implies uncertainties when applying Eq. (8) in the main text to systems with a 
positive $Q$-value in which the hindrance is not observed experimentally.

\begin{figure}[htbp]
  \centering
  \includegraphics[width=8.6cm]{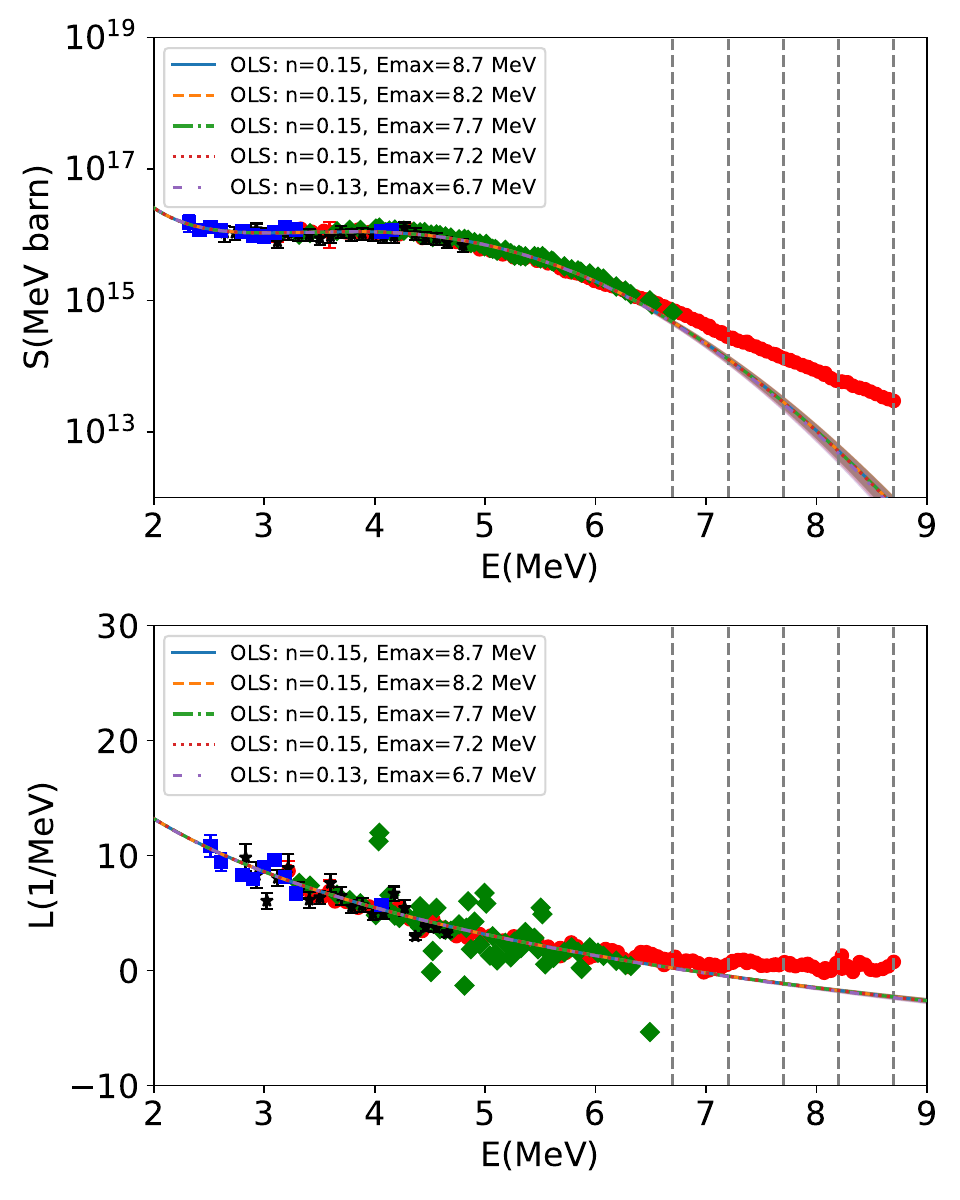} 
  \caption{Same as Fig. 6, but for the OLS fitting. 
  }
  \label{fig8}
\end{figure}

In the case of the OLS fitting,
the relative weight of the higher-energy data points is
much smaller than in the OLS fitting. 
Because of this, 
the uncertainty originating from the energy cut-off becomes much smaller, as is demonstrated in 
Fig. \ref{fig8}. 
We point out that our conclusion of the absence of the hindrance phenomenon in 
$^{12}$C+$^{13}$C does not change with different choices of the cut-off energy (See Fig. 9). 

\begin{figure}[htbp]
  \centering
  \includegraphics[width=8.6cm]{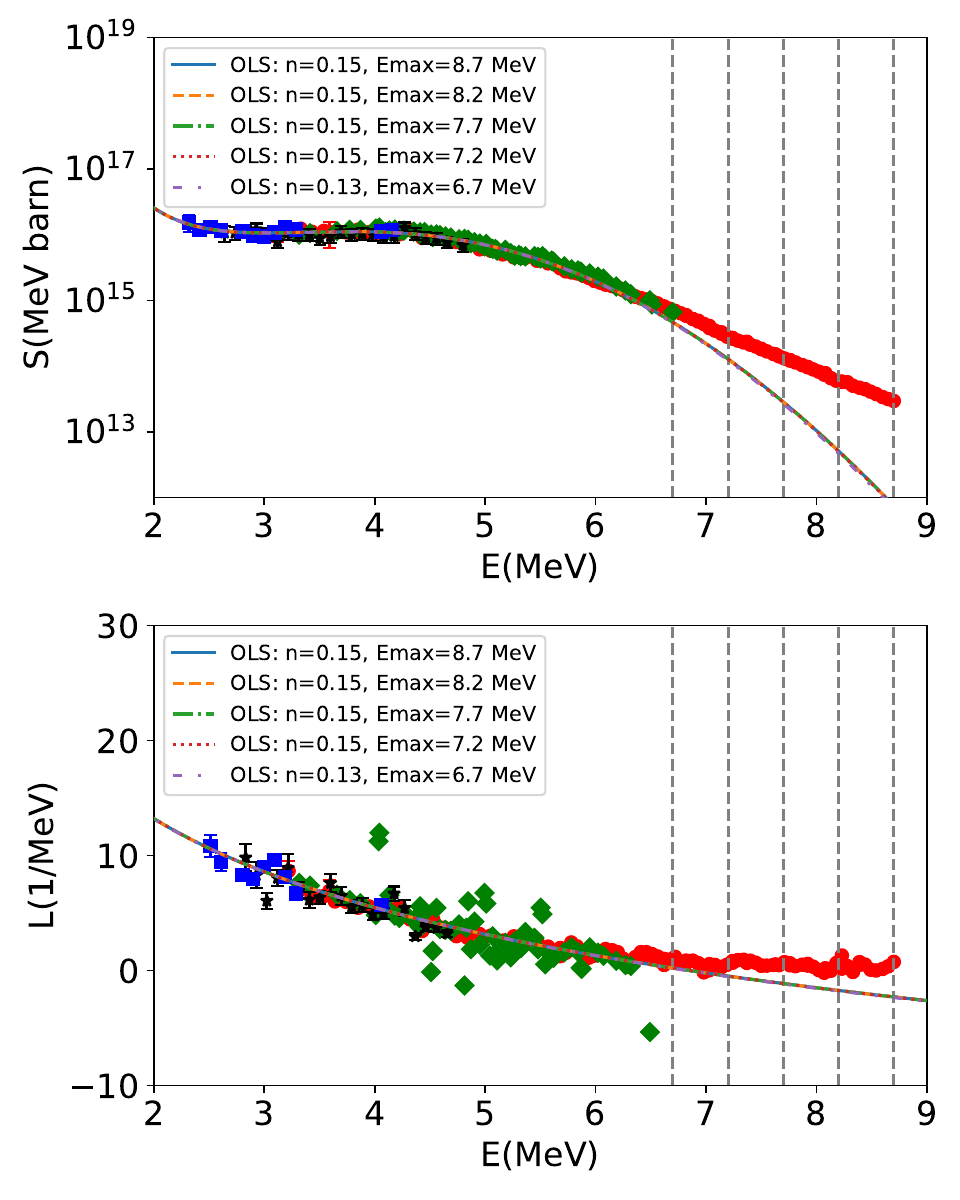} 
  \caption{Same as Fig. 7, but for the OLS fitting. 
  }
  \label{fig9}
\end{figure}